# Bounds for the Sum Capacity of Binary CDMA Systems in Presence of Near-Far Effect

P. Pad, M. H. Shafinia, S. M. Mansouri, P. Kabir, F. Marvasti

Advanced Communications Research Institute (ACRI),

Department of Electrical Engineering, Sharif University of Technology, Tehran, Iran

Email: {pedram_pad, shafinia, m_mansouri, pouryakabir}@ee.sharif.edu and marvasti@sharif.edu

*Abstract*—In this paper we are going to estimate the sum capacity of a binary CDMA system in presence of the near-far effect. We model the near-far effect as a random variable that is multiplied by the users binary data before entering the noisy channel. We will find a lower bound and a conjectured upper bound for the sum capacity in this situation. All the derivations are in the asymptotic case. Simulations show that especially the lower bound is very tight for typical values $E_b/N_0$ and near-far effect. Also, we exploit our idea in conjunction with the Tanaka's formula [6] which also estimates the sum capacity of binary CDMA systems with perfect power control.

## I. INTRODUCTION

In a DS-CDMA system each user is assigned a signature vector used to transmit data through a common channel. Each user multiplies its signature by its data and transmits it through the channel. All the vectors are added up together in the channel and the resultant vector embedded in noise is received at the receiver end. In practical situations, it is preferred to use binary antipodal signatures in conjunction with BPSK. Thus, the channel model is

$$Y = \frac{1}{\sqrt{m}}AX + N \quad (1)$$

where $\frac{1}{\sqrt{m}}A$ is the $m \times n$ signature matrix with entries $\pm\frac{1}{\sqrt{m}}$ ($n$ is the number of users and $m$ is the chip rate) and $X$ is the users data vector with entries belonging to $\{\pm 1\}$. $N$ is the noise vector whose entries are $i.i.d$ with PDF $f$.

Without perfect power control the assumption of receiving equal powers from all senders is no more valid. Therefore, the channel changes to a more real one:

$$Y = \frac{1}{\sqrt{m}}AP^{½}X + N \quad (2)$$

where $P^{½} = \text{diag}(\sqrt{p_1}, \cdots, \sqrt{p_n})$ in which $p_1, \cdots, p_n$ are the users power receives at the receiver.

The near-far effect has a stochastic nature but since our systems have some power control mechanisms (may be non-perfect) we can suppose that it has values around unity. Consequently, we model it as a random variable with mean $+1$. If the PDF of each $\sqrt{p_i}$ is $\tilde{g}(x)$ which is symmetric around $+1$, we can convert the previous model to the following equivalent one

$$Y = \frac{1}{\sqrt{m}}A(X + Z) + N \quad (3)$$

in which $Z$ is a vector with independent entries with PDF $g(x) = \tilde{g}(x + 1)$. Clearly, $g(x)$ is a symmetric PDF around 0.

In our previous work [1], we supposed that entries of $Z$ are $i.i.d$ random variables with uniform distribution in the interval $[-\eta, \eta]$ and found lower and upper bounds on $\eta$ for which the errorless communication is possible and proposed methods for constructing large size signature matrices which guarantee errorless communication for moderate values of $\eta$. In addition, we suggested almost ML decoder for a subclass of proposed near-far resistant codes.

Most of the other works done on this topic concentrate on decoding methods such as MMSE in combination with Successive Interference Cancellation (SIC) [2] and blind adaptive interference suppression [3]. In [4] authors suggest a method called Isolation Bit Insertion (IBI) for overcoming the near-far problem. Additionally, authors of [5] have found lower and upper bounds for near-far resistance of MMSE detector.

Additionally, some works have been done about estimating the sum capacity of binary CDMA systems. In [6] an estimation of the sum capacity when the system scale tends to infinity and $n/m \to \beta$, for a given $\beta$, is proposed. The authors of [7] that the formula derived in [6], is an upper bound for the sum capacity of binary CDMA systems. In [8-10] a lower bound and a conjectured upper bound are proposed for the sum capacity of finite dimensional binary CDMA systems. The behavior of these bounds in the limiting case is also derived in [8]. Both [8] and [9] are based on averaging the mutual information of the input ($X$) and the output ($Y$) over all signature matrices.

In this paper, using the theorems about the distribution of the eigen values of random matrices, we are going to estimate the sum capacity of CDMA system in presence of near-far effect in the asymptotic case by exploiting the results derived in both [6] and [8].

In the next section, we state the basic theorems which we will use for deriving our results. Sections III and IV contain deriving the lower and upper bounds, respectively. In section V, some simulation results for the lower and upper bounds are done. Summary and future works are covered in section VI.

## II. PRELIMINARIES

In this paper we are going to estimate

$$\frac{1}{n}C(m,n,f,g) = \max_{A \in \mathcal{M}_{n \times m}(\{\pm 1\})} \max_{p(x_1)\ldots p(x_n)} \mathbb{I}(X;Y)$$

where $Y$ and $X$ has the relation stated in Eq. (3) ($f$ and $g$ are the PDF of the entries of $N$ and $Z$, respectively) as $n, m \to \infty$ and $n/m \to \beta$ (for a given $\beta$). We denote this quantity by $c(\beta, f, g)$, i.e.,

$$c(\beta, f, g) = \lim_{\substack{n,m \to \infty \\ n/m \to \beta}} \frac{1}{n} C(m,n,f,g).$$

We concentrate on the case that $f$ is a Gaussian random variable with zero mean and variance $\sigma^2$ and $g$ is a Gaussian random variable with zero mean and variance $\rho^2$ and use $c(\beta, \sigma, \rho)$ instead of $c(\beta, f, g)$. For the case $\beta \leq 1$, the orthogonal codes are optimum and the system performance is equivalent to Binary PSK. Thus,

$$c(\beta, \sigma, \rho) = h(\widehat{w}) - h(w)$$

where $w$ is the PDF of a Gaussian random variable with zero mean and variance $\sigma^2 + \rho^2$,

$$\widehat{w}(x) = \frac{w(x-1) + w(x+1)}{2}$$

and $h$ is the differential entropy. Consequently, the open problem is $c(\beta, \sigma, \rho)$ for $\beta > 1$. Notice that in overloaded CDMA system ($\beta > 1$), the power estimation methods at the receiver fails and the receiver cannot have a good estimate of the users power. Thus, in overloaded CDMA systems, the receiver has not noticeable knowledge about the users powers.

In [8], lower and upper bounds were derived for $c(\beta, \sigma, 0)$ (perfect power control). In fact the upper bound is based on a conjecture that has not been proven yet. It assumes that the uniform distribution on the users binary data maximizes the mutual information for any signature matrix. The bounds are as follows.

**Lower Bound [8]** For any $\gamma$,

$$c(\beta, \sigma, 0) \geq 1 - \inf_{\gamma} \sup_{t \in [0,1]} \left[ H(t) + \frac{1}{2\beta}\left(\gamma \log e - \log\left(1 + \gamma\left(1 + \frac{4t\beta}{\sigma^2}\right)\right)\right)\right]$$

**Conjectured Upper Bound [8]**

$$c(\beta, \sigma, 0) \leq \min\left\{1, \frac{1}{2\beta}\log\left(1 + \frac{\beta}{\sigma^2}\right)\right\}.$$

According to the claim of authors of [7], the formula derived in [6] is an upper bound for $c(\beta, \sigma, 0)$. Thus, we use it to derive another upper bound for $c(\beta, \sigma, \rho)$. The formula in [6] is

**Tanaka Upper Bound [6]**

$$C_m = \frac{1}{2\beta}\log\left(1 + \frac{\beta(1-m)}{\sigma^2}\right) + g(\lambda, m)\log e$$

in which

$$g(\lambda, m) = \frac{\lambda}{2}(1+m) - \int \ln(\cosh(\sqrt{\lambda}Z + \lambda))D_Z$$

where $D_Z$ is the standard normal measure and $\lambda = \frac{1}{\sigma^2 + \beta(1-m)}$ and $m = \int \tanh(\sqrt{\lambda}Z + \lambda)D_Z$.

We will use the following theorem in our approach of estimating $c(\beta, \sigma, \rho)$.

**Marčenko-Pastur Theorem [10]** Consider an $N \times K$ matrix $H$ whose entries are independent zero-mean complex (or real) random variables with variance $1/N$ and fourth moments of order $O(1/N^2)$, as $K, N \to \infty$ with $K/N \to \beta$, the empirical distribution of $H^T H$ converges almost surely to a nonrandom limiting distribution with density

$$f_\beta(x) = \left(1 - \frac{1}{\beta}\right)^+ \delta(x) + \frac{\sqrt{(x-a)^+(b-x)^+}}{2\pi\beta x}$$

where $a = (1 - \sqrt{\beta})^2$, $b = (1 + \sqrt{\beta})^2$ and $(Z)^+ = \max(Z, 0)$.

Now, we are ready to start.

## III. LOWER BOUND

In this section we are going to derive lower bound for $c(\beta, \sigma, \rho)$.

**Theorem 1** For any $\gamma$ we have

$$c(\beta, \sigma, \rho) \geq 1 - \inf_{\gamma} \sup_{t \in [0,1]} \left[ H(t) + \frac{1}{2\beta}\left(\gamma \log e - \log\left(1 + \gamma\left(1 + \frac{4t\beta}{\theta^2}\right)\right)\right)\right]$$

where $\theta^2 = (\sqrt{\beta} + 1)^2 \rho^2 + \sigma^2$.

*Proof*: According to Eq. (3),

$$Y = \frac{1}{\sqrt{m}}A(X + Z) + N = \frac{1}{\sqrt{m}}AX + \frac{1}{\sqrt{m}}AZ + N$$
$$= \frac{1}{\sqrt{m}}AX + \widetilde{N} \qquad (4)$$

where $\widetilde{N} = \widetilde{Z} + N$ and $\widetilde{Z} = \frac{1}{\sqrt{m}}AZ$. Since $Z = [Z_1, \ldots, Z_m]^T$ and $Z_i$'s are $i.i.d$ random variables with the distribution $\mathcal{N}(0, \rho^2)$, $\frac{1}{\sqrt{m}}AZ$ is a zero mean Gaussian vector with covariance matrix $\frac{\rho^2}{m}AA^T$. Let $H = \frac{1}{\sqrt{n}}A^T$, according to Marčenko-Pastur theorem, the emprical distribution of $\frac{1}{n}AA^T$ is less than $\left(1+\frac{1}{\sqrt{\beta}}\right)^2 \rho^2$ with probability 1, as $m, n \to \infty$. We know that the covariance matrix of $\widetilde{N}$ is $\frac{\rho^2}{m}AA^T + \sigma^2 I$. According to above discussion, the maximum eigen value of $\frac{\rho^2}{m}AA^T + \sigma^2 I$ is $\left(\sqrt{\beta}+1\right)^2 \rho^2 + \sigma^2$. Therefore, if we substitute $\widetilde{N}$ in (4) by a Gaussian random vector with $i.i.d.$ entries with zero mean and variance $\left(\sqrt{\beta}+1\right)^2 \rho^2 + \sigma^2$, we arrive at a system with lower sum capacity. Eventually, combining the above point with the lower bound stated in the Preliminaries, the proof completes. ∎

In fact the lower bound is obtained by substituting the maximum probable eigen value of $\frac{\rho^2}{m}AA^T + \sigma^2 I$ instead of the variance of noise $\sigma^2$ in the lower bound formula.

## IV. UPPER BOUND

This section will be devoted to finding upper bound for the sum channel capacity in the presence of near-far effect.

**Theorem 2**

$$c(\beta, \sigma, \rho) \leq \min\left\{h(\hat{g}) - h(g), \frac{1}{2\beta}\log\left(1 + \frac{\beta}{\omega^2}\right)\right\}$$

where $\omega^2 = \left(\sqrt{\beta}-1\right)^2 \rho^2 + \sigma^2$.

*Proof*: The first term in the min function is an obvious upper bound because according to data-processing inequality $\mathbb{I}(X;Y) \leq \mathbb{I}(X;X+Z)$ and the maximum of the right-hand side of this inequality is $n\bigl(h(\hat{g}) - h(g)\bigr)$.

Now, the second term: According to the discussion in the proof of Theorem 1 and the Marčenko-Pastur theorem, the eigen values of $\frac{\rho^2}{m}AA^T + \sigma^2 I$ is greater than $\left(\sqrt{\beta}-1\right)^2 \rho^2 + \sigma^2$, with probability 1 as $m, n \to \infty$. Now, if we substitute $\widetilde{N}$ with a noise that all of its eigen values are $\left(\sqrt{\beta}-1\right)^2 \rho^2 + \sigma^2$ we arrive at a system with higher capacity. Since all of the eigen values of the covariance matrix of the noise vector are equal, their entries are $i.i.d.$ (it is not hard to prove). Combining this points with the conjectured upper bound stated in the Preliminaries, the proof completes.∎

Also, we can obtain another upper bound by substituting $\left(\sqrt{\beta}-1\right)^2 \rho^2 + \sigma^2$ instead of $\sigma^2$ in the Tanaka's upper bound. We will simulate all of these formulas for various values of $\beta$, $\rho$ and $\sigma$ in the next section.

**Note 1** According to the simulations in [8] and what stated in [10] about the Marčenko-Pastur theorem, the asymptotic analysis that is performed here is very close to finite case for $m, n \geq 32$. Thus, the lower and upper bounds derived here can be assumed as lower and upper bounds for the sum capacity of CDMA systems with usual sizes.

**Note 2** Since for a given variance the Gaussian distribution has the maximum entropy, the lower bound derived in this paper is a lower bound for the sum capacity of any system with the variance of near-far equal $\rho^2$.

## V. SIMULATION RESULTS

The simulations are done based on a parameter we call it power control factor (PCF). We define:

$$PCF_{dB} = 10\log_{10}\frac{\mathbb{E}(\tilde{g})^2}{\text{Var}(\tilde{g})}$$

PCF is an index that shows the amount of the power fluctuations of the users. Following simulations show the behavior of the bounds derived in this paper.

In the first case, we simulated lower bounds for an overloaded binary CDMA system with $\beta = 2$ and several values for PCF. This simulation shows the effect of PCF on the channel capacity and as it can be predicted greater PCF results greater capacity. The second simulation shows the conjectured upper bound for different values of PCF when $\beta = 4$.

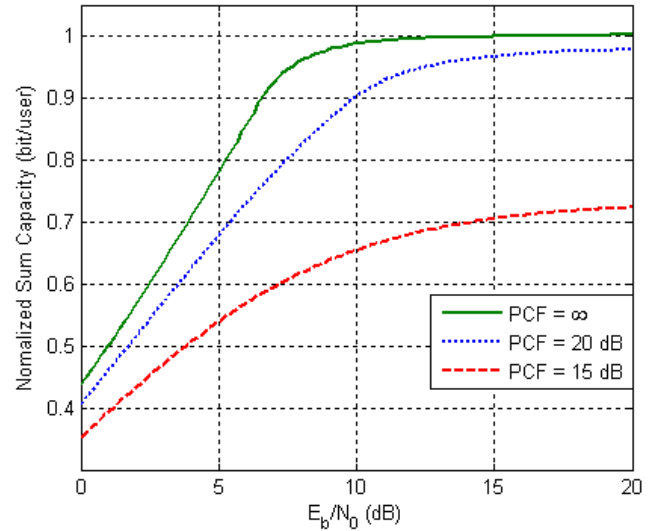

Fig. 1. The normalized sum capacity lower bounds vs. $E_b/N_0$ for $\beta = 2$ and various values of PCF.

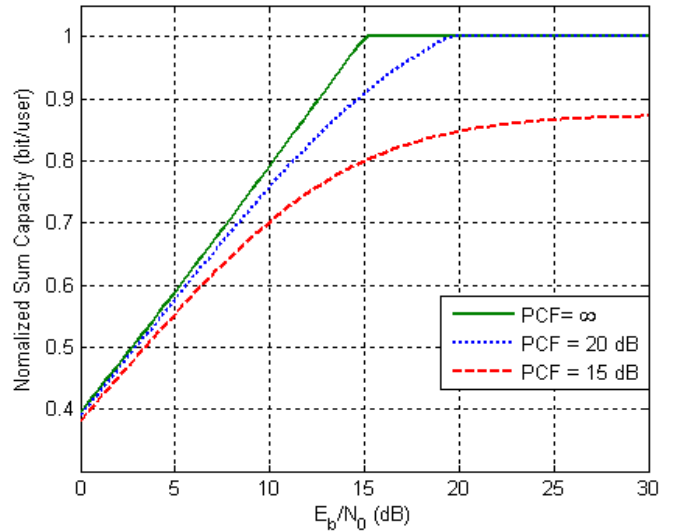

Fig. 2. The conjectured upper bounds for normalized sum capacity vs. $E_b/N_0$ for $\beta = 4$ and different values of PCF.

Figures 3 and 4 show a comparison of the proposed bounds including the one of Tanak for $\beta = 2$ and 4, respectively. Thsese simulations show that the systems with higher overloading factor (higher ratio of number of users to the chip rate) will be more damaged with non-perfect power control.

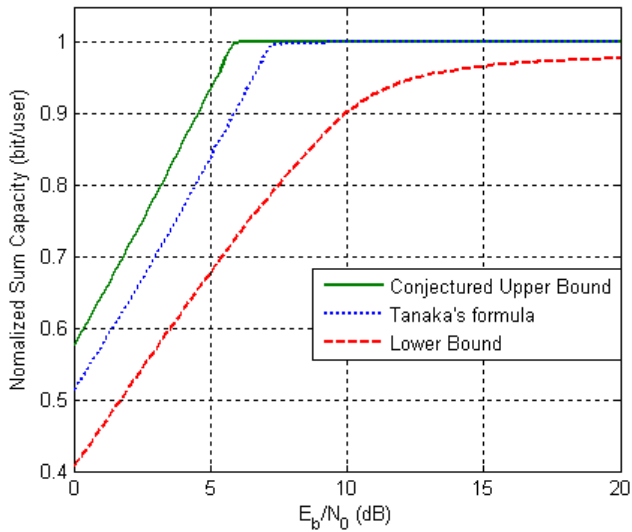

Fig. 3. The lower and upper bounds for the normalized sum capacity vs. $E_b/N_0$ for $\beta = 2$ and PCF = 20 dB.

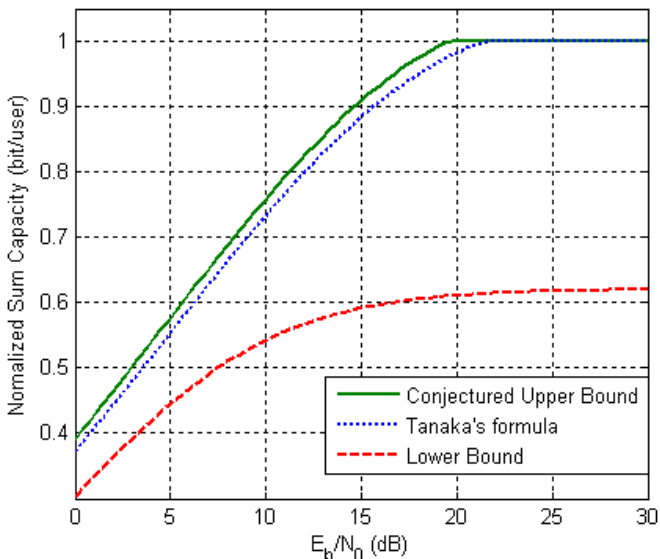

Fig. 4. The lower and upper bounds for the normalized sum capacity vs. $E_b/N_0$ for $\beta = 4$ and PCF = 20 dB.

Figure.5 shows another comparison of our lower and upper bounds with Tanaka's bounds as PCF varies, assuming that $\beta = 2$ and $E_b/N_0 = 20$. This simulation shows that CDMA systems with PCF greater than 35 dB perform very close to perfectly power controlled systems. Also, it shows that PCF lower than 20 dB certainly degrades the system performance.

## VI. CONCLUSION

In this paper some lower and upper bounds are derived for the sum capacity of binary CDMA systems in the presence of near-far effect. Tending the power control factor to infinity, we arrive at the previously known bounds for the sum capacity of binary CDMA system. There are two points about the current work that can be interesting ideas for future works. The first one is that the near-far effect only affects the power of the received signal. Thus, in the case of using BPSK, the sign of users data does not change due to near-far effect. Consequently, the system model used here is worse than what occurs in practice because in our model it is probable that $\text{sign}(X_i + Z_i) \neq \text{sign}(X_i)$.

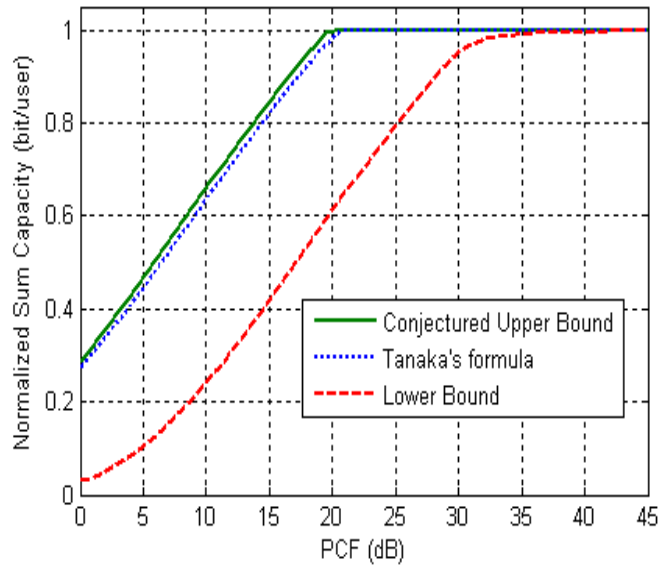

Fig. 5. Lower and upper bounds for the normalized sum channel capacity vs. PCF for $\beta = 2$ and $E\_b/N\_0 = 20$.

The second point is that, the near-far effect of each user is assumed to be $i.i.d.$ random variables at different times in the current model. But in practical situations, the time samples of the near-far effect of each user change continually and thus are not independent from each other. Thus, the case studied in this paper is a lower bound for the performance of the practical systems. Assuming the channel with memory makes the problem much more complex to solve but will gives more practical bounds. The current work is good for estimating the performance of systems with high fluctuations in users power.

## REFERENCES

[1] M. H. Shafinia, P. Kabir, P. Pad, S. M. Mansouri, F. Marvasti, "Errorless codes for CDMA system with near-far effect," *Accepted in ICC'10*.

[2] Y. Cho and J. H. Lee, "Analysis of an Adaptive SIC for Near-Far Resistant DS-CDMA," *IEEE Transactions on Communications*, vol. 46, No. 11, pp. 1429-1432, 1998.

[3] U. Madhow, "Blind Adaptive Interference Suppression for the Near-Far Resistant Acquisition and Demodulation of Direct-Sequence CDMA Signals," *IEEE Transactions on Signal Processing*, vol. 45, no. 1, pp. 124-136, 1997.

[4] F. C. Zheng, S. K. Barton, "Near-Far Resistant Detection of CDMA Signals via Isolation Bit Insertion," *IEEE Transactions on Communications*, vol. 43, no. 2/3/4, pp. 1313-1317, 1995.

[5] U. Madhow and M. L. Honig, "On the Average Near-Far Resistance for MMSE Detection of Direct Sequence CDMA Signals with Random Spreading," *IEEE Transactions on Information Theory*, vol. 45, no. 6, pp. 2039-2045, 1999.

[6] T. Tanaka, "A statistical-mechanics approach to large-system analysis of CDMA multiuser," *IEEE Trans. Inform. Theory*, vol. 48, no. 11, pp. 2888–2910, Nov. 2002.

[7] S. B. Korada and N. Macris, "Tight bounds on the capacity of binary input random CDMA systems," *ArXiv: 0803.1454v1 [cs.IT]*, Mar. 2008.

[8] K. Alishahi, F. Marvasti, V. Aref, and P. Pad, "Bounds on the Sum Capacity of Synchronous Binary CDMA Channels," *IEEE Trans. On Inform. Theory*, vol. 55, no. 8, pp. 3577-3593, Aug. 2009.

[9] P. Pad, F. Marvasti, K. Alishahi, and S. Akbari, "A class of errorless codes for over-loaded synchronous wireless and optical CDMA systems," *IEEE Trans. on Inform. Theory*, vol. 55, no. 6, pp. 2705-2715, June 2009.

[10] P. Pad, F. Marvasti, K. Alishahi, and S. Akbari, "Errorless codes forover-loaded synchronous CDMA systems and evaluation of channel capacity bounds", *in Proc. of Int. Symp. Information Theory* (*ISIT'08*), Toronto, Canada, June 2008.

[11] A. M. Tulino, S.Verdu ,*Random Matrix Theory and Wireless Communications* ,  now Publisher Inc., 2004.